%% file: main.tex
\documentclass[conference]{IEEEtran}
\usepackage{graphicx}
\usepackage{booktabs}
\usepackage{amsmath,amssymb}
\usepackage{url}
\usepackage{cite}
\usepackage{tikz}
\usetikzlibrary{arrows.meta,positioning,calc}
\usepackage{comment}
\usepackage{xcolor}
\usepackage{bbm} % Added for proper rendering of the indicator function
\usepackage{bbm}
\newcommand{\remove}[1]{}
\newcommand{\figSize}[0]{0.50}
\newcommand{\figSizeBig}[0]{0.60}

\begin{document}
\title{ITS Fairy: Occlusion Assistance Selected Against a Recipient's Own Perception Reports}
\author{
\IEEEauthorblockN{Yenan Wang, Oscar Karlsson, and Elad M. Schiller}
\IEEEauthorblockA{Department of Computer Science and Engineering\\
Chalmers University of Technology\\
Gothenburg, Sweden\\
\{yenan, oscakarl, elad\}@chalmers.se}
\and
\IEEEauthorblockN{Francesco Raviglione\IEEEauthorrefmark{1} and Claudio Casetti\IEEEauthorrefmark{2}}
\IEEEauthorblockA{\IEEEauthorrefmark{1}Department of Electronics and Telecommunications\\
\IEEEauthorrefmark{2}Department of Control and Computer Engineering\\
Politecnico di Torino\\
Turin, Italy\\
\{francesco.raviglione, claudio.casetti\}@polito.it}
}
\maketitle

\input{manuscript.tex}

\bibliographystyle{IEEEtran}
\bibliography{references}
\end{document}

%% file: manuscript.tex
% VTC 2027 Spring -- ITS Fairy compressed manuscript draft
% Scope: sensing-occluded lane merge and four-way intersection only.

\begin{abstract}
Cooperative perception can expose object state beyond a vehicle's onboard sensors, but sensing occlusion can still leave a local safety application without the objects its collision computation needs. To tackle this challenge, we present the ITS Fairy, an infrastructure-side Server Local Dynamic Map (S-LDM) service whose decision unit is the pair \emph{(recipient, missing conflict-relevant object)}: among objects absent from a recipient's CPM-derived reported awareness, it sends only those relevant to a Time of Closest Approach (TCA) conflict test. Comparable services predict what a vehicle can perceive; the ITS Fairy instead reads what it has already reported. The recipient inserts the selected state into its local LDM and uses its unchanged collision-avoidance controller. We evaluate this application-level mechanism in SUMO--ms-van3t--S-LDM emulation, since extended as VaN3Twin, using a sensing-occluded lane merge and four-way intersection scenario. At every main-sweep speed, the smallest assisted per-encounter minimum TCA exceeds the largest local-only value in the archived data.
Additionally, assisted medians remain in the multi-second range where local-only operation repeatedly approaches zero. In the lane-merge robustness data, the median benefit persists at 80\% configured assistance omission with 10 and 5~Hz analysis, but largely disappears at 100--120~km/h when 80\% omission is combined with 1~Hz analysis. These results demonstrate the application-level value of supplying object state selected against what a recipient has itself reported. They are not a vehicular wireless-channel evaluation, and they do not quantify what selectivity saves relative to forwarding every nearby object.
\end{abstract}

\begin{IEEEkeywords}
cooperative perception, infrastructure-assisted perception, server local dynamic map, sensing occlusion, collision avoidance
\end{IEEEkeywords}

\section{Introduction}
\label{sec:introduction}

A collision-avoidance application can act only on the traffic participants
represented in its local environment model. Sensing occlusion can hide exactly
the object that a conflict computation needs: two vehicles approaching an
obstructed junction may be on intersecting paths while neither appears in the
other's Local Dynamic Map (LDM). Cooperative perception and an infrastructure-hosted Server
Local Dynamic Map (S-LDM)~\cite{sldm} may nonetheless hold that object state, aggregated
from the CAMs and CPMs that multiple vehicles already
%
%transmit~\cite{etsi_en_302637-2,etsi_ts_103324}. Providing everything the infrastructure holds to every vehicle is wasteful and overutilizes the wireless channel; hence, a recipient-specific question arises: of the object state available at the infrastructure, which part should be supplied to \emph{this} vehicle? 
%
transmit~\cite{etsi_en_302637-2,etsi_ts_103324}. Providing all of it to every vehicle is wasteful of the shared wireless channel, so a recipient-specific question arises: which part should be supplied to \emph{this} vehicle?

%Prior work differs less in whether it answers that question than in what evidence it uses to answer it.

\smallskip
\noindent \textbf{Open Issues and Positioning.~~}
Infrastructure-assisted perception and its operational constraints are already
well studied, including end-to-end delay in cloud-assisted
pipelines~\cite{Hawlader2024VTC}, roadside sensor fusion~\cite{Pinho2024VTC},
communication delay in context-aware collision avoidance~\cite{Goetz2024VTC},
information freshness~\cite{Huang2024VTC}, receiver-side reduction of redundant
CPM processing~\cite{Karunathilake2025VTC}, and a real obstructed intersection
with active traffic-moderating infrastructure~\cite{Khoshkdahan2026VTC}. What
remains open is not whether infrastructure can assist, but how the information
it provides is selected.

The simplest selection rules consult only the sender's own object list and the
state of the channel. The Collective Perception Service
(CPS)~\cite{etsi_ts_103324} filters through generation rules and redundancy
mitigation applied to the sender's own detections; RRS~\cite{rrs} ranks objects
by an accident-risk indicator and by how often each has already been reported;
and AICP~\cite{aicp} sorts objects by informativeness to limit channel load and
on-screen clutter. These rules establish what is worth sending in general. However, none
of them determines what a particular vehicle is missing to enable collision avoidance effectively.

Another family of rules relies on predicted risk rather than on the recipient's
perceptual state. Ma \emph{et al.}~\cite{Ma2026RICPN} decide whether cooperation
is warranted for a road user, from collision risk and blind-spot
occupancy; SRA-CP~\cite{Liu2025SRACP} has vehicles exchange coverage summaries
and select peers for risk-relevant blind zones; and STDR-CP~\cite{stdrcp}
aggregates per-vehicle demand into a global broadcast schedule.
The closest related work, IBCP~\cite{ibcp}, conditions on an ego's declared intent and
returns conflict alerts aimed only at the road users relevant to that ego's
announced maneuver, evaluated with a minimum-margin metric under occlusion. Risk
and intent identify which encounters matter, but not which participant lacks the object state that makes the encounter dangerous.

Closer to this work are approaches that \emph{estimate} what a particular
recipient already holds. AutoCast~\cite{autocast} computes, for every object and
neighbour, whether the object is geometrically visible from that neighbour's
pose and whether it is relevant to the neighbour's broadcast trajectory.
CCPAV~\cite{ccpav} scores an object for a specific receiver as its unseen-ness
divided by the time until the two trajectories intersect, and arbitrates
centrally under a resource constraint. Lv \emph{et al.}~\cite{lv} learn
perceptibility from object geometry and occlusion, and disseminate only the
objects predicted to be unseen. Higuchi \emph{et al.}~\cite{higuchi} maintain,
for each neighbour, an anticipated history of the messages it is likely to have
received, and include a perception record only when it would measurably change
that neighbour's inferred belief. Abdel-Aziz \emph{et al.}~\cite{abdelaziz} pair
vehicles at the infrastructure and let each sender learn which regions to share
from its partner's feedback. In all of them the recipient's awareness is
inferred rather than observed, and each estimator carries a cost: geometric
visibility presupposes a model of the recipient's sensors, learned
perceptibility is weakest where several vehicles converge, and feedback-driven
schemes require a return channel that Higuchi \emph{et al.}~\cite{higuchi}
judged too expensive in vehicle-to-vehicle operation to justify the savings. A parallel line of work makes related decisions on sparse feature or point-based
representations inside a multi-agent model~\cite{where2comm,coopernaut} rather
than on object state. 

Table~\ref{tab:positioning} summarizes this progression,
from no recipient model, through estimated recipient state, to what the
recipient has reported. The open issue is therefore this: recipient-specific
supplementation needs to rest on evidence of what the recipient has reported
perceiving, and to send only those missing objects that bear on an actual
conflict.

\begin{table}[t]
	\centering
	\caption{\label{tab:positioning}Work, awareness source, and resulting action.}
	\scriptsize
	\setlength{\tabcolsep}{2.0pt}
	\begin{tabular}{@{}p{0.245\columnwidth}p{0.36\columnwidth}p{0.36\columnwidth}@{}}
		\toprule
		Work & How awareness is obtained & Resulting action \\
		\midrule
		RRS~\cite{rrs} & Not modelled; object redundancy counted across received CPMs. & Fixed-size object set in a broadcast CPM. \\
		AICP~\cite{aicp} & Not modelled; objects sorted by informativeness. & Filters objects for forwarding and driver display. \\
		Ma \emph{et al.}~\cite{Ma2026RICPN} & Not modelled; collision risk and blind-spot occupancy. & Identifies CP necessity from infrastructure observations. \\
		IBCP~\cite{ibcp} & The ego's declared intent. & Issues conflict alerts into a dedicated braking pathway. \\
		Lv \emph{et al.}~\cite{lv} & Estimated: learned perceptibility from object geometry. & Disseminates objects predicted imperceptible. \\
		AutoCast~\cite{autocast} & Estimated: geometric visibility from the neighbour's pose. & Schedules point-cloud transmission between vehicles. \\
		CCPAV~\cite{ccpav} & Estimated: receiver line-of-sight, scaled by time to trajectory intersection. & Prioritises centrally relayed messages. \\
		Higuchi \emph{et al.}~\cite{higuchi} & Estimated: anticipated message history per neighbour. & Includes records that change an inferred belief. \\
		Abdel-Aziz \emph{et al.}~\cite{abdelaziz} & Estimated: learned from the paired receiver's feedback. & Transmits selected regions to the paired vehicle. \\
		ITS Fairy & Read: the recipient's own reported CPM history. & Sends object state into the recipient LDM; maneuver remains local. \\
		\bottomrule
	\end{tabular}
\end{table}

%Abdel-Aziz \emph{et al.}~\cite{abdelaziz} pair vehicles at the infrastructure and let each sender learn, which regions to share from the satisfaction feedback its partner returns. What these approaches share is that the recipient's awareness is inferred. None of them directly observes whether the recipient has itself reported the object, and each estimator carries its own cost: geometric visibility presupposes a model of the recipient's sensors, learned perceptibility reports its weakest accuracy where several vehicles converge, and feedback-driven schemes require a return channel that Higuchi \emph{et al.}~\cite{higuchi} judged too expensive in vehicle-to-vehicle operation to justify the savings. Table~\ref{tab:positioning} summarizes this progression, from selection that does not model the recipient, through selection that estimates the recipient's state, to selection based on what the recipient has reported.

%The open issue is therefore specific. An infrastructure service needs a basis
%for recipient-specific supplementation that rests on %evidence of what the
%recipient has itself reported perceiving, and it needs %to restrict what it sends
%to those missing objects that bear on an actual conflict.

\smallskip \noindent \textbf{Proposed Approach.~~}
We propose a novel design concept, named \emph{ITS Fairy}, whose 
%
%to address this issue directly. Its 
%
decision unit is the pair \emph{(recipient, missing conflict-relevant object)}. For each vehicle, the service derives a recipient-specific reported-awareness set from that vehicle's own CPM history, and forms the residual between the fresh S-LDM objects in the vehicle's neighbourhood and the objects the vehicle has itself reported. A conflict test based on Time of Closest Approach (TCA), an established criticality metric for automated driving~\cite{westhofen}, is applied to that residual alone; for each surviving object the Fairy transmits that object's state, and neither a collision verdict nor a maneuver command. The recipient inserts the supplied state into its local LDM, recomputes TCA, and acts through its unchanged avoidance controller. Westhofen \emph{et al.} refer to this quantity as Time To Closest Encounter; we use the term TCA throughout.

This is affordable for architectural rather than algorithmic reasons: an S-LDM already ingests the CPMs of every vehicle in its coverage area, so the awareness set is a byproduct of aggregation and requires no additional message, which is why the option dismissed as too costly between vehicles is available here. Holding the local controller unchanged is likewise deliberate, so that an observed difference is attributable to information availability rather than to a different predictor.

%What makes the reported-awareness model affordable is architectural rather than
%algorithmic. An S-LDM already ingests the CPMs of every vehicle in its coverage area, hence
%the awareness set is a byproduct of aggregation and requires no message that is
%not already being sent. This is precisely why the option was dismissed in
%the vehicle-to-vehicle setting and it becomes available here. Holding the local
%controller unchanged is likewise a deliberate part of the experimental design:
%it allows an observed difference to be attributed to the availability of
%information rather than to a different predictor or controller.

\smallskip \noindent \textbf{Evaluation.~}
Two questions drive our evaluation. 
\textbf{RQ1:} under sensing occlusion, how
does recipient-specific missing-object assistance change the minimum-TCA margin
available to an unchanged local collision-avoidance pipeline, relative to
local-only operation? 
This is examined in a sensing-occluded lane merge and four-way intersection, with local-only and assisted operation sharing the same local controller, using per-encounter minimum TCA as the primary metric and route traversal time as secondary.
\textbf{RQ2:} in the lane-merge, how does the retained minimum-TCA
benefit vary with configured assistance-message omission and with the Fairy's analysis rate? 
Omission is swept from 0 to 80\%, and the Fairy's analysis rate, \emph{i.e.}, how often it recomputes a vehicle's missing-object residual and tests it for conflict, is set to 10, 5, and 1~Hz.
That delivery loss, staleness and interruption matter is established rather than claimed here: object state has been delivered into an unmodified automated-driving stack on real vehicles at an occluded intersection~\cite{asabe}, and the effects of unreliable and interrupted communication studied directly~\cite{thunberg,ren}. RQ2 asks the narrower question of how this mechanism degrades, not how a wireless channel behaves.

%Delivery loss, staleness and connectivity interruption matter for
%infrastructure-assisted perception and this is established rather than claimed here:
%object state has been delivered into an unmodified automated-driving stack on
%real vehicles at an occluded intersection, with the stack performing the
%avoidance maneuver itself~\cite{asabe}, and the effects of unreliable
%communication and interrupted cooperation have already been studied
%directly~\cite{thunberg,ren}. RQ2 asks the narrower, application-level question
%of how this particular mechanism degrades in presence of the conditions mentioned earlier, and not how a wireless channel behaves.

\smallskip \noindent \textbf{Findings.~~}
We claim that supplying conflict-relevant object state, selected against what a recipient has itself reported, improves the information available to an unchanged local safety controller under the tested occlusion conditions.

\emph{Effectiveness.~~} Assistance separates local-only and Fairy-assisted operation completely in the obtained results: at every evaluated speed in both geometries, the smallest
assisted per-encounter minimum TCA exceeds the largest local-only value. Assisted medians hold at multi-second values where local-only medians sit at or below 0.08~s over most of the sweep, and the four-way intersection reproduces the effect in a multi-vehicle geometry. Lane-merge traversal time also falls, by about 11\% to 45--47\% across 30--120~km/h.

%\emph{Effectiveness.~~} Assistance separates the local-only and Fairy-assisted configurations completely in the obtained results: at every evaluated speed in both geometries, the smallest
%assisted per-encounter minimum TCA exceeds the largest local-only value. In the
%lane merge, local-only medians are 0.46~s at 30~km/h and at most 0.08~s from
%40--120~km/h, whereas assisted medians hold at 7.47--9.47~s from 40--70~km/h and
%decline to 3.83~s at 120~km/h. The four-way intersection reproduces the effect
%in a multi-vehicle geometry: we achieved 1.25~s local-only against 4.83~s assisted at
%30~km/h, and local medians of 0.01--0.03~s from 40--80~km/h against assisted
%medians declining from 4.65 to 2.24~s. Lane-merge traversal time also falls,
%with the mean reduction growing from about 11\% to 45--47\% across 30--120~km/h.

\emph{Robustness.~~} The benefit is jointly conditioned on omission and analysis
frequency rather than on omission alone. At 10~Hz the benefit survives 80\% omission almost intact; at 1~Hz the same omission leaves it only at the lowest tested speed, and the collapse is abrupt rather than gradual.

%The benefit is jointly conditioned on omission and analysis frequency rather than on omission alone.
%At 10~Hz and 80\% configured omission, median minimum TCA is still 5.49, 4.68 and 3.61~s at 90, 100 and 120~km/h, against local-only medians of 0.01, 0.04 and 0.08~s. At 1~Hz the same omission leaves 2.29~s at 90~km/h but 0.04 and 0.08~s at 100 and 120~km/h, essentially matching local-only operation. The 1~Hz distributions are visibly bimodal, so the median shifts abruptly rather than degrading smoothly.

\smallskip \noindent \textbf{Contributions.~~}
\label{sec:intro-contributions}
The ITS Fairy addresses a challenge created by sensing occlusion, which can leave a vehicle's local safety application without the state of a conflict-relevant object that is nevertheless available at the infrastructure. The Fairy uses the recipient's own CPM-derived reported awareness to identify such missing objects, filters them for conflict relevance, and supplies their state while collision assessment and maneuver selection remain local. Our contributions are:

\begin{itemize}
    \item A novel infrastructure-side awareness model that reads recipient awareness from the recipient's own CPM reports rather than inferring it, and uses it to identify and supply conflict-relevant missing object state.

    \item An open-source SUMO--ms-van3t--S-LDM evaluation environment implementing the ITS Fairy
    and the sensing-occluded lane-merge and four-way-intersection scenarios.

    \item An evaluation showing substantial minimum-TCA benefits, and a joint characterization of how they depend on assistance omission and analysis rate.
    
    %, which to our knowledge has not been reported for infrastructure-assisted perception.
    
\end{itemize}

\noindent We pledge to release the evaluation environment and the archived result data upon acceptance; Section~\ref{sec:methodology} states which experimental configurations are preserved in the repository. We expect the ITS Fairy concept, its recipient-specific awareness-repair method, and the accompanying evaluation environment to provide an open basis for studying targeted infrastructure assistance and to inform future cooperative-perception designs.

\section{System Model and ITS Fairy}
\label{sec:system-model}
 
\begin{figure}[t]
\centering
\resizebox{0.97\columnwidth}{!}{%
\begin{tikzpicture}[
    >=Stealth,
    every node/.style={font=\scriptsize}
]
 
  \node[
    draw,rounded corners,align=center,
    minimum width=12mm,minimum height=9mm
  ] (hidden) {Occluded\\object};
 
  \node[
    draw,rounded corners,align=center,
    minimum width=16mm,minimum height=13mm,
    right=28mm of hidden
  ] (fairy) {S-LDM\\+ ITS Fairy};
 
  \node[
    draw,rounded corners,align=center,
    minimum width=16mm,minimum height=13mm,
    right=30mm of fairy
  ] (recv) {Recipient\\local LDM};
 
  \node[
    draw,rounded corners,align=center,
    minimum width=16mm,minimum height=8mm,
    below=5mm of recv
  ] (ctrl) {local TCA + maneuver};
 
  % Occluded object -> infrastructure
  \draw[->,thick]
    (hidden.east) --
    node[above,align=center]{CAM/CPM\\to infrastructure}
    (fairy.west);
 
  % Recipient -> Fairy
  \draw[->,thick]
    (recv.west) to[bend right=12]
    node[above,align=center]{CAM/CPM\\reports}
    (fairy.east);
 
  % Fairy -> Recipient
  \draw[->,thick]
    (fairy.east) to[bend right=12]
    node[below,align=center]{selected missing\\object state}
    (recv.west);
 
  \draw[->,thick]
    (recv) -- (ctrl);
 
  \draw[dashed,thick]
    (hidden.south) to[bend right=25]
    node[below]{blocked sensing LOS}
    (recv.south);
 
\end{tikzpicture}%
}
\caption{\label{fig:architecture}ITS Fairy information flow. The S-LDM aggregates CAM/CPM information, while the Fairy selectively supplies object state missing from a recipient's reported awareness. Collision-risk recomputation and maneuver control remain local.}
\end{figure}
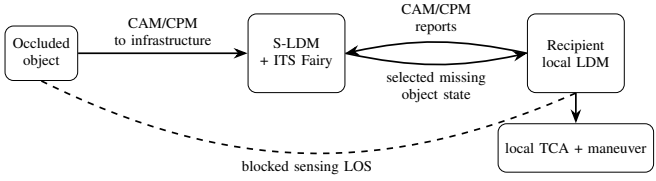

\noindent \textbf{System Model.~~}
Vehicles maintain local LDMs and generate CAMs describing their state and CPMs describing perceived objects~\cite{etsi_en_302637-2,etsi_ts_103324}. An
infrastructure node receives these reports, maintains a centralized S-LDM, and hosts the Fairy~\cite{sldm}. Onboard sensing is range-limited and can be obstructed, while infrastructure connectivity is assumed available and is modelled at the application level. The Fairy has no privileged view of the environment: its only input is a CAM/CPM-derived S-LDM state describing connected and sensed road users, which is what makes the recipient's reported awareness the quantity the service can act on, rather than its actual perception. In the experiments, the simulator's ground truth is indeed never provided. Objects are associated by vehicle identifier and stale entries are handled by the S-LDM. In our implementation, a local LDM cleaner runs every 0.5~s and removes entries older than 1~s; Fairy-supplied objects use the same local-LDM aging path after insertion. Fig.~\ref{fig:architecture} summarizes the information flow and the separation between infrastructure-side information selection and vehicle-local maneuver control.  
 
 % \remove
\noindent \textbf{The ITS Fairy.~~}%
Fig.~\ref{fig:algorithm} states the procedure executed by the ITS Fairy, and its line labels are cited below. For recipient $v$, let $\mathcal{B}_v(t)$ be fresh S-LDM vehicle objects within the implementation's dynamic neighborhood, whose radius is $\max\{\|\mathbf v_v\|T_h,d_{\rm base}\}$ with prediction horizon $T_h=10$~s and minimum neighborhood radius $d_{\rm base}=5$~m, following~\cite{malinverno}. Let $\mathcal{D}_v(t)$ be the S-LDM history of objects reported in $v$'s CPMs. The proof-of-concept assumes this reported set corresponds to the objects in $v$'s local LDM, yielding the missing set (F1--F3)
\begin{equation}
\mathcal{M}_v(t)=\mathcal{B}_v(t)\setminus\mathcal{D}_v(t).
\label{eq:missing-set}
\end{equation}
For each $o\in\mathcal{M}_v(t)$, the Fairy applies a 2-D constant-velocity TCA relevance calculation (F4) with a 10~s prediction horizon. With relative position $\mathbf r$ and velocity let $\mathbf u$, $t^*=\mathrm{clip}(-\mathbf r\!\cdot\!\mathbf u/(\mathbf u\!\cdot\!\mathbf u),0,T_h)$, with the zero-relative-velocity case handled separately; hence, the predicted separation at that instant is $d^{*}=\|\mathbf r+\mathbf u\,t^{*}\|$. An object is selected when its predicted separation at $t^*$ is below a fixed Fairy relevance threshold $d_F=5$~m (F5) (independent of $d_{\rm base}$), so that $\operatorname{Risk}_{\rm TCA}(v,o,t)=\mathbbm{1}[d^{*}<d_F]$, giving
\begin{equation}
\mathcal{A}_v(t)=\{o\in\mathcal{M}_v(t):\operatorname{Risk}_{\rm TCA}(v,o,t)=1\}.
\label{eq:assist-set}
\end{equation}
For each selected object (F6), the Fairy sends one custom CAM-like message carrying the recipient identifier, object station identifier, position/elevation, heading, speed, timestamps, dimensions, and station type; it sends no acceleration, confidence, or TCA verdict. The receiver derives local identifier and relative pose, inserts the state into its LDM, and recomputes TCA (R1--R2). 
The unchanged local controller also uses constant-velocity TCA over 10~s, but applies it to every object in its local LDM rather than to a speed-dependent neighborhood.
%
%The unchanged local controller also uses constant-velocity TCA over 10~s, but with additional prefilters and a vehicle-dimension-based separation threshold. 
%
One message is sent per selected object, without acknowledgments/retransmissions, in line with how CAMs and CPMs are normally disseminated \cite{etsi_en_302637-2,etsi_ts_103324}.

\begin{figure}[t]
\centering
%\scriptsize
\footnotesize
\setlength{\tabcolsep}{2.5pt}
\begin{tabular}{@{}r l@{}}
\multicolumn{2}{@{}l}{\textbf{Fairy on updated recipient $v$}}\\
F1:& $\mathcal{B}\leftarrow$ fresh S-LDM neighbors of $v$\\
F2:& $\mathcal{D}\leftarrow$ objects in $v$'s CPM-derived history\\
F3:& \textbf{for each} $o\in\mathcal{B}\setminus\mathcal{D}$ \textbf{do}\\
F4:& \quad $(t^*,d^*)\leftarrow\operatorname{TCA}(v,o)$\\
F5:& \quad \textbf{if} $d^*<d_F=5$~m \textbf{then}\\
F6:& \qquad send $\operatorname{ObjectState}(o)$ to $v$\\[0.5mm]
\multicolumn{2}{@{}l}{\textbf{Recipient on ObjectState$(o)$}}\\
R1:& map $o$ to a local id and pose, insert into the local LDM\\
R2:& recompute local TCA and apply the unchanged maneuver policy\\
\end{tabular}
\caption{Selective missing-object assistance. F-steps run at the infrastructure, R-steps at the recipient. The Fairy decides what state to supplement; collision-risk recomputation and control remain at the recipient.}
\label{fig:algorithm}
\end{figure}

\section{Experimental Methodology}
\label{sec:methodology}
 
\noindent \textbf{Setup and Measurement.~~}%
This section describes how we evaluate RQ1 and RQ2. The same comparison answers both questions: an unchanged vehicle-local collision-avoidance mechanism, operating on its local LDM, is run with and without Fairy assistance. %inserted into that LDM. 
We compare local-only and Fairy-assisted operation in the same SUMO--ms-van3t--S-LDM co-simulation~\cite{msvan3t,sldm}. ms-van3t~\cite{msvan3t}, since extended as VaN3Twin~\cite{van3twin}, generates and encodes CAMs/CPMs, but the evaluated path carries them over UDP/IPv4 and virtual Ethernet to a UDP-to-AMQP relayer and then to the S-LDM, which is subscribed to an AMQP broker \cite{sldm}; the ITS Fairy uses the same broker to return the state. No vehicular wireless PHY/MAC or propagation model is instantiated, and vehicle applications deliberately do not consume direct peer V2V messages. The robustness parameter is therefore an independent application-level omission applied immediately before Fairy-state insertion into the local LDM, not a measured or simulated wireless Packet Delivery Ratio (PDR). This isolates the mechanism's dependence on assistance availability from the propagation and congestion effects that cause loss in deployment; it does not approximate correlated wireless loss.
 
The primary metric is the per-encounter minimum TCA, \emph{i.e.}, the smallest TCA value the vehicle-local detector produces during an encounter, considering the vehicles participating in that encounter. It is only defined once the detector holds both objects, so it measures the margin available after the conflict becomes locally computable. A vehicle-local \texttt{determineRisk()} function runs every 100~ms; a separate AMQP thread subscribes to a Fairy-message topic, the simulator polls it every 10~ms, and successful insertion triggers immediate TCA reevaluation. This rate is fixed across all conditions and independent of the Fairy analysis rate swept in RQ2, where the Fairy queries the S-LDM for updated stations every 100/200/1000~ms for 10/5/1~Hz. Route traversal time from spawn to completion is a secondary metric. 

\noindent \textbf{Experimental Design.~~}%
We consider two main scenarios: lane merge and four-way intersection.
The lane merge considers equal-length lanes separated by a building that blocks the 50~m vehicle onboard sensing LOS; the building affects perception, but not the communication path to the infrastructure. Vehicles reach the merging point approximately together; the closer one has priority, with a tie-break based on vehicle identifiers.  The four-way scenario considers four simultaneously spawned straight-through vehicles at 30--80~km/h. Buildings obstruct sensing, vehicles yield to traffic on the right, and lower IDs break the four-way tie; route-dependent traversal times are reported separately. Fig.~\ref{fig:scenarios} shows both scenarios and road geometries. The nominal sweep is 30--120~km/h and 140--200~km/h is an extreme-speed stress regime.
 
\begin{figure}[t]
\centering
\begin{minipage}[t]{0.52\textwidth}
\centering
\begin{tikzpicture}[scale=0.45]
\definecolor{roadgray}{RGB}{100,100,100}
\def\halfw{0.4}
\begin{scope}[rotate=-15]
  \fill[roadgray] (-9,\halfw) rectangle (0,-\halfw);
  \draw[white,line width=1pt] (-9,\halfw)--(0,\halfw);
\end{scope}
\begin{scope}[rotate=15]
  \fill[roadgray] (-9,\halfw) rectangle (0,-\halfw);
  \draw[white,line width=1pt] (-9,-\halfw)--(0,-\halfw);
\end{scope}
\fill[roadgray] (0,\halfw) rectangle (6,-\halfw);
\fill[brown!70] (-1.8,0)--(-8.75,1.8)--(-8.75,-1.8)--cycle;
\begin{scope}[shift={(-5.5,1.47)},rotate=-15]
 \fill[red!70] (-0.45,-0.22) rectangle (0.45,0.22);
 \draw[->,thick,red,dashed] (0.45,0)--(5.75,0);
\end{scope}
\begin{scope}[shift={(-5.5,-1.47)},rotate=15]
 \fill[blue!70] (-0.45,-0.22) rectangle (0.45,0.22);
 \draw[->,thick,blue,dashed] (0.45,0)--(5.75,0);
\end{scope}
\end{tikzpicture}\\[-1mm]
\small (a) Lane merge with blocked sensing LOS.
\end{minipage}
\begin{minipage}[t]{0.44\textwidth}
\centering
\begin{tikzpicture}[scale=0.30,>=Stealth]
\definecolor{roadgray}{RGB}{100,100,100}
\fill[roadgray] (-1,6) rectangle (1,-6);
\fill[roadgray] (-6,1) rectangle (6,-1);
\fill[brown!70] (1.25,-1.25) rectangle (6,-6);
\fill[brown!70] (1.25,1.25) rectangle (6,6);
\fill[brown!70] (-1.25,1.25) rectangle (-6,6);
\fill[brown!70] (-1.25,-1.25) rectangle (-6,-6);
\fill[red!70] (0.25,-4) rectangle (0.75,-3.2);
\fill[yellow!70] (-0.25,4) rectangle (-0.75,3.2);
\fill[blue!70] (4,0.75) rectangle (3.2,0.25);
\fill[green!70] (-4,-0.75) rectangle (-3.2,-0.25);
\draw[->,thick,red,dashed] (0.5,-3.2)--(0.5,4);
\draw[->,thick,yellow,dashed] (-0.5,3.2)--(-0.5,-4);
\draw[->,thick,blue,dashed] (3.2,0.5)--(-4,0.5);
\draw[->,thick,green,dashed] (-3.2,-0.5)--(4,-0.5);
\end{tikzpicture}\\[-1mm]
\small (b) Four-way intersection with four occluded approaches.
\end{minipage}
\caption{Evaluated sensing-occluded geometries. Buildings obstruct onboard sensing while infrastructure connectivity is retained.}
\label{fig:scenarios}
\end{figure}
 
This design has known limitations that bound how the results should be read.
The route generators schedule 100 sequential encounters per speed and configuration within one simulation execution: two vehicles per lane-merge encounter and four per intersection encounter. Routes and nominal speed are fixed, and our implementation sets a SUMO seed equal to 10, hence these are not independently seeded replications; therefore, we emphasize descriptive distributions. All configurations contain 100 encounters except the 10~Hz, 120~km/h cells at 40\% and 70\%, with 99 and 92 usable encounters. SUMO collision reports serve only as supporting evidence, because simulator hitboxes differ from the controller geometry.
% PROVENANCE -- UNVERIFIED. Confirm against the archive before submission.
% Current text asserts these files underlie the original thesis evaluation.
% If that cannot be established, state only which files produced the reported
% numbers and make no claim about the alternates.
% These archived primary files are those underlying the original thesis/report evaluation, and reproduce its corresponding figure.
 
The robustness tests at 90, 100, and 120~km/h drop each Fairy-to-vehicle assistance message independently at the configured omission rate, leaving CAM/CPM reporting to the S-LDM unaffected.
With 10~Hz Fairy analysis, the omission rate takes the values 0, 5, 10, 15, 20, 30, 40, 50, 60, 70, and 80\%; the 5~Hz and 1~Hz sweeps use the same values except 15\%.
Table~\ref{tab:experimental-matrix} lists the full configuration set.
 
\begin{table}[t]
\centering
\caption{Experimental matrix. Frequencies denote the Fairy analysis rate, not the CAM/CPM generation frequency.}
\label{tab:experimental-matrix}
\scriptsize
\setlength{\tabcolsep}{2.6pt}
\begin{tabular}{@{}llll@{}}
\toprule
Experiment & Speeds (km/h) & Assistance omission & Frequency \\
\midrule
Lane merge & 30--120 & 0\% & 10~Hz \\
Stress regime & 140--200 & 0\% & 10~Hz \\
Robustness & 90/100/120 & 0--80\% & 10/5/1~Hz \\
Four-way & 30--80 & 0\% & 10~Hz \\
\bottomrule
\end{tabular}
\end{table}

\section{Results}
\label{sec:results}

\begin{figure*}[t]
\centering
\begin{minipage}[t]{\figSizeBig\textwidth}
\centering
\includegraphics[width=\linewidth]{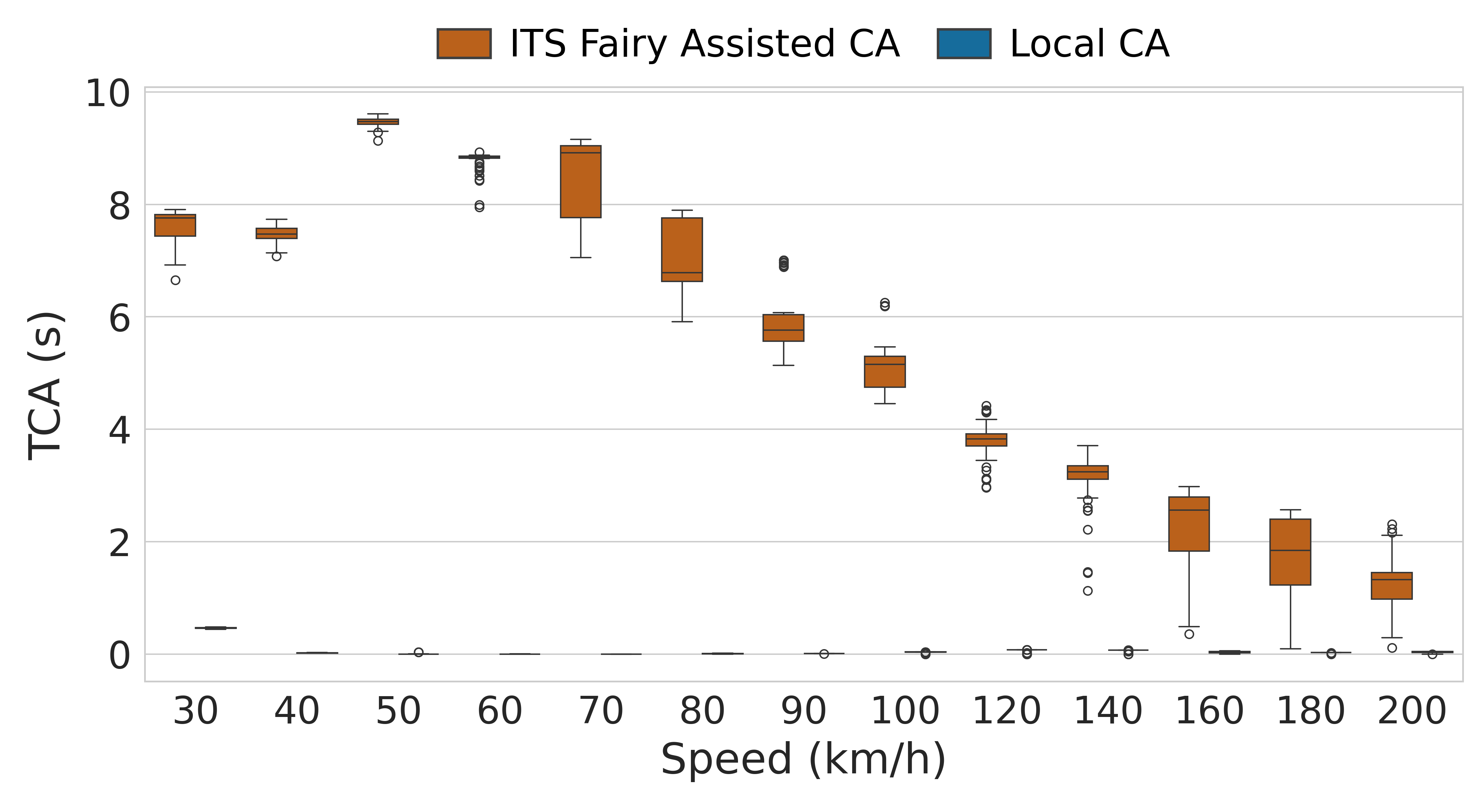}\\[-1mm]
\small (a) Sensing-occluded lane merge.
\end{minipage}\hfill
\begin{minipage}[t]{\figSizeBig\textwidth}
\centering
\includegraphics[width=\linewidth]{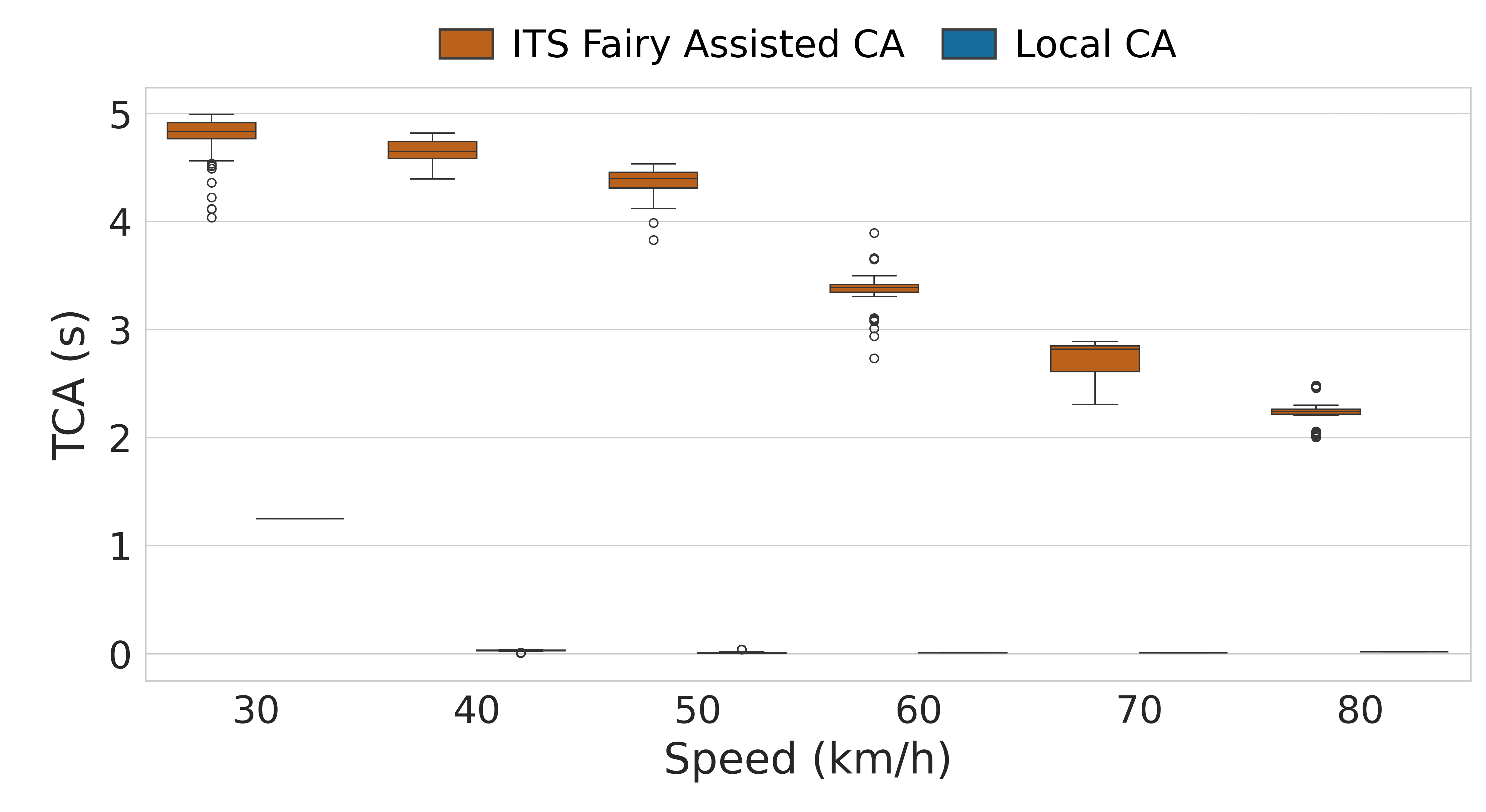}\\[-1mm]
\small (b) Sensing-occluded four-way intersection.
\end{minipage}
\caption{Per-encounter minimum TCA for local-only and Fairy-assisted operation (100 sequential encounters per speed and configuration in the main sweeps); the axis title abbreviates this as TCA. Local-only boxes are flattened near zero at most speeds. Speeds above 120~km/h in (a) are an extreme-speed stress regime.}
\label{fig:primary-results}
\end{figure*}

\smallskip \noindent \textbf{Effect of missing-object assistance.~~}
Fig.~\ref{fig:primary-results} shows the primary minimum-TCA results. In the lane merge, the local-only median is 0.46~s at 30~km/h and at most 0.08~s from 40--120~km/h. Fairy-assisted medians remain 7.47--9.47~s from 40--70~km/h, then decrease to 6.78, 5.76, 5.15, and 3.83~s at 80, 90, 100, and 120~km/h. The stress regime identifies the limit rather than a deployment range: assisted medians decline to 3.24, 2.56, 1.85, and 1.33~s at 140, 160, 180, and 200~km/h. Across every evaluated lane-merge speed, the smallest assisted minimum-TCA value exceeds the largest local-only value in the obtained results. Lane-merge traversal time is also lower over the nominal sweep: from 30 to 120~km/h, the mean reduction grows from about 11\% to 45--47\% on the two equal-length routes. Our analysis attributes this to the avoidance style rather than to higher speeds being sustained: local-only vehicles detect the conflict late and brake hard, frequently to a full stop, whereas assisted vehicles adjust speed gradually. Because vehicles were configured to continue after stopping, local-only traversal times also include encounters that ended in a simulator collision. Above 140~km/h the traversal-time difference narrows; we report this as a coincident trend rather than attributing it to the TCA change.

The four-way intersection shows the same effect in a multi-vehicle geometry. At 30~km/h, the local-only median minimum TCA is 1.25~s versus 4.83~s with assistance. From 40--80~km/h, local medians are only 0.01--0.03~s, which is deemed too low to effectively avoid collisions.
Indeed, actuation phase delays on the order of a few tenths of a second have been recently measured in experimental automated vehicles \cite{werner2026driftingfuturestabilizingpath}.
Conversely, assisted medians decrease gradually from 4.65 to 2.24~s. Moreover, the \emph{lowest} assisted minimum TCA among all 100 sequential encounters is still 4.39, 3.83, 2.73, 2.31, and 2.00~s at 40, 50, 60, 70, and 80~km/h, respectively; at every vehicle speed, the assisted value also exceeds the largest local-only value. Additionally, route-specific traversal times generally improve. Across the three routes other than Route~4, mean traversal-time reductions span 12.3--50.2\% over 30--80~km/h. We observed that Route~4 shows a smaller reduction and reverses at 80~km/h (19.78~s assisted versus 18.75~s local) ; because the precise cause of this reversal was not disentangled, we report it without attributing it to the Fairy.

\begin{figure*}[t]
\centering
\begin{minipage}[t]{\figSize\textwidth}
%\begin{minipage}[t]{0.46\textwidth}
\centering
\includegraphics[width=\linewidth]{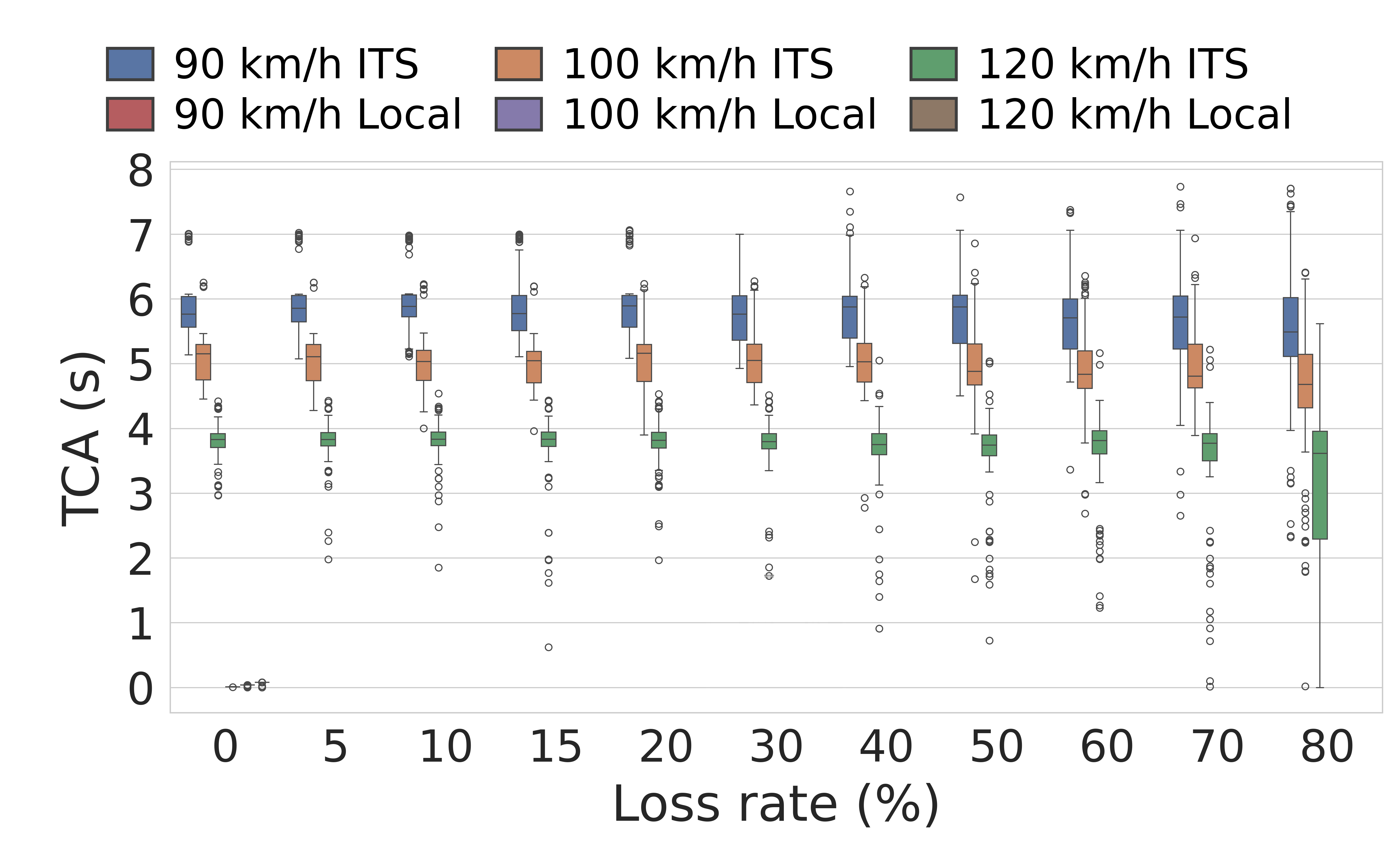}\\[-1mm]
\small (a) 10~Hz
\end{minipage}\hfill
\begin{minipage}[t]{\figSize\textwidth}
%\begin{minipage}[t]{0.46\textwidth}
\centering
\includegraphics[width=\linewidth]{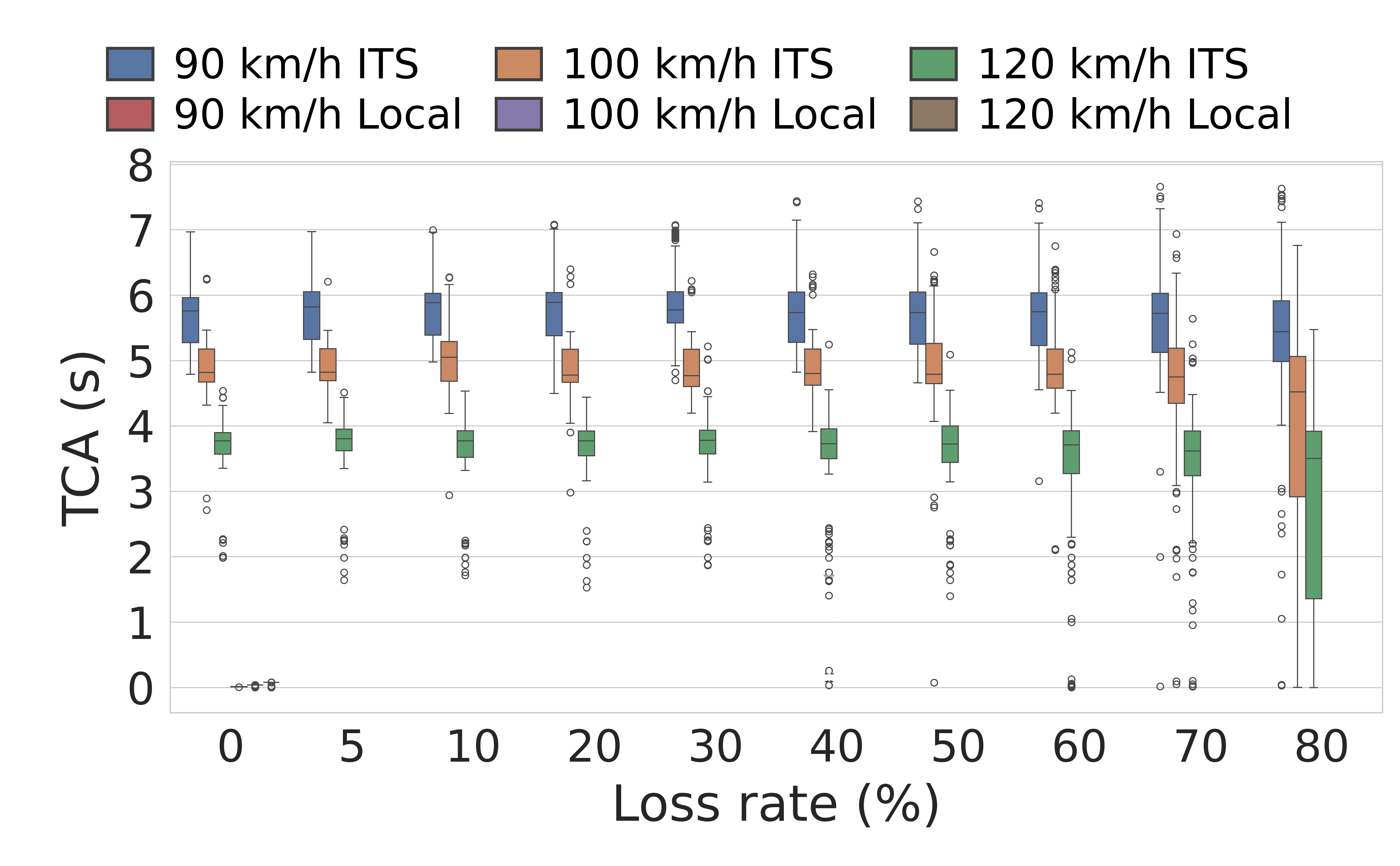}\\[-1mm]
\small (b) 5~Hz
\end{minipage}\\[1mm]
\begin{minipage}[t]{\figSize\textwidth}
%\begin{minipage}[t]{0.46\textwidth}
\centering
\includegraphics[width=\linewidth]{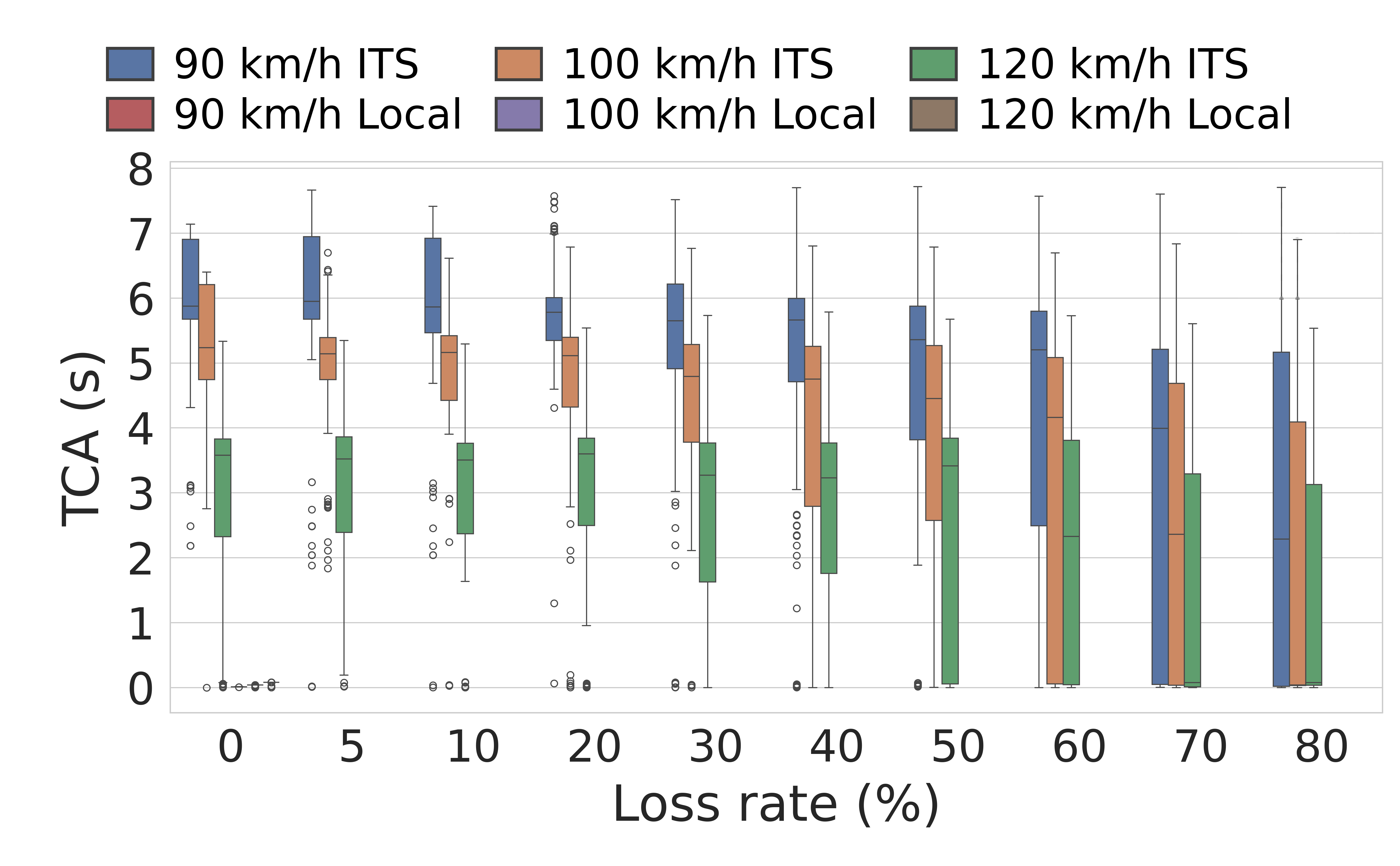}\\[-1mm]
\small (c) 1~Hz
\end{minipage}
\caption{Lane-merge per-encounter minimum-TCA distributions across the archived Fairy-to-vehicle assistance-omission sweeps at 10, 5, and 1~Hz analysis frequency. The local-only distributions are omission-independent and are drawn once, at the 0\% position, as a reference. The 15\% cell is absent at 5 and 1~Hz. At 1~Hz and high omission the median line coincides with the lower box edge; Table~\ref{tab:robustness-80} gives those values numerically.}
\label{fig:robustness-sweep}
\end{figure*}

\smallskip \noindent \textbf{Robustness to missed assistance opportunities.~~}
Fig.~\ref{fig:robustness-sweep} shows the results for the assistance-omission sweeps. At 10~Hz, the median minimum TCA changes modestly as the configured independent application-level assistance omission increases. Even at 80\% omission, the median minimum TCA is 5.49, 4.68, and 3.61~s at 90, 100, and 120~km/h respectively, versus local-only medians of 0.01, 0.04, and 0.08~s. At 5~Hz the corresponding medians remain similar, but the lower tail worsens; at 120~km/h the first quartile falls from 2.29~s to 1.36~s. At 1~Hz the omission sensitivity is much stronger: with 80\% omission, the median remains 2.29~s at 90~km/h but falls to 0.04 and 0.08~s at 100 and 120~km/h, essentially matching local-only operation (Table~\ref{tab:robustness-80}). The 1~Hz distributions are also visibly bimodal: individual encounters either retain most of the assisted margin or collapse onto the local-only value, so the median shifts abruptly rather than degrading smoothly and understates the spread between the two groups. For these three tested speeds, the median benefit therefore persists even at 80\% configured omission with 10--5~Hz analysis, whereas the 1~Hz data show greater sensitivity to missed assistance opportunities. Quartiles characterize this bimodal regime only partially. It is worth noting how these configured independent omissions characterize application-level sensitivity, and not a tolerable packet loss rate for a real wireless link.

%substantially

\begin{table}[t]
\centering
\caption{Median minimum TCA at 80\% Fairy-to-vehicle assistance omission; first quartile in parentheses.}
\label{tab:robustness-80}
\scriptsize
\setlength{\tabcolsep}{3.0pt}
\begin{tabular}{@{}lccc@{}}
\toprule
 & 90 & 100 & 120~km/h \\
\midrule
Fairy 10~Hz & 5.49 (5.11) & 4.68 (4.32) & 3.61 (2.29) \\
Fairy 5~Hz  & 5.44 (4.99) & 4.52 (2.92) & 3.51 (1.36) \\
Fairy 1~Hz  & 2.29 (0.02) & 0.04 (0.03) & 0.08 (0.03) \\
Local-only  & 0.01 (0.01) & 0.04 (0.04) & 0.08 (0.08) \\
\bottomrule
\end{tabular}
\end{table}

\smallskip
\noindent \textbf{Interpretation.~~}
\emph{RQ1.~~}The comparison supports an information-availability reading rather
than a better-predictor reading. Recipient-side prediction, thresholds, maneuver
logic and vehicle dynamics are unchanged; the assisted vehicle differs only in
receiving selected S-LDM object state and recomputing TCA locally. The shape of
the separation supports this. At every evaluated lane-merge speed the smallest
assisted minimum TCA exceeds the largest local-only value, so the two
distributions are disjoint rather than shifted: a difference in predictor quality
would be expected to move a distribution, not to separate it. The assisted medians
also decline smoothly with speed, from 7.47--9.47~s at 40--70~km/h to 3.83~s at
120~km/h, while local-only medians stay pinned at or below 0.08~s across
40--120~km/h. A margin that erodes gradually with closing speed while the
unassisted baseline remains flat is the signature of earlier information, not of
a different decision rule.
The metric must be read with its construction in mind. Minimum TCA comes from the same local
detector that drives the controller, and under occlusion that detector cannot evaluate a conflict
until the other vehicle enters the 50~m sensing model, which is why local-only values sit near
zero. Part of the separation therefore reflects \emph{when} the conflict became locally computable, the deficit the Fairy removes, so local-only values are not an independent measure of
physical proximity, and a near-zero value does not distinguish a collision from a vehicle that stopped just in time.
The minimum TCA is an operational conflict-imminence measure: the comparison gives evidence about
information availability and the margin left to the controller, and it is not a collision-rate result.

\emph{RQ2.~}The frequency--omission interaction follows from how much road is
covered between successive analyses. At 10~Hz the Fairy re-examines a recipient
every 100~ms; at 1~Hz once per second,
which at 120~km/h corresponds to roughly 33~m of distance travelled against a 50~m sensing
model. The observed behavior matches the obtained results. At 10~Hz the benefit survives
80\% omission almost intact, with medians of 5.49, 4.68 and 3.61~s at 90, 100 and
120~km/h against local-only medians of 0.01, 0.04 and 0.08~s; even at 80\% omission a recipient still receives about two assistance opportunities per second on average, so one typically arrives before the conflict becomes critical. At 1~Hz the same omission leaves
2.29~s at 90~km/h but 0.04 and 0.08~s at 100 and 120~km/h, and the collapse is
abrupt rather than gradual. The 1~Hz distributions are correspondingly bimodal:
individual encounters either retain most of the assisted margin or fall back to
the local-only value. This bimodality is what a delivery-opportunity mechanism predicts: an object either reaches the recipient before the conflict becomes critical or contributes almost nothing.

\section{Conclusion}
\label{sec:conclusion}
The ITS Fairy turns a broader centralized view, based on a Server Local Dynamic Map (S-LDM), into recipient-specific assistance by
sending only conflict-relevant object state absent from a vehicle's reported
awareness, leaving collision-risk recomputation and maneuver control local.
Reading that awareness from the recipient's own CPMs, rather than inferring it
from geometry or predicted perceptibility, costs no additional message, besides the ones an aggregation point
is already receiving. In sensing-occluded lane merging, assisted minimum TCA
holds at 3.83--9.47~s where the same controller without assistance stays at or
below 0.08~s from 40--120~km/h, and mean traversal time falls by about 11\% to
45--47\% across that range. The benefit survives 80\% assistance omission at
10~Hz analysis but not at 1~Hz, indicating that what matters is delivery
opportunity rather than delivery volume. Standardized messaging, measured delay
and age of information, and controlled-seed replication are the natural next
steps.

%\smallskip
%\noindent \textbf{Limitations.~~}
%
%Three limits bound what the archived data can settle. The evaluation has no
%complementary baseline in which the infrastructure forwards every nearby object
%irrespective of the recipient's reported awareness, so selectivity is established
%as a property of the decision rule rather than as a measured saving, and the
%experiments cannot show whether the same TCA margins would follow from unfiltered
%forwarding at higher and unquantified cost. Physical closest-separation traces
%were not preserved, so contact and stopped-in-time encounters cannot be separated
%from the archive. The robustness omission draws were not seeded, so those cells
%are not reproducible draw-for-draw, and no exact historical revision of the
%experimental configuration survives; the current repository is the closest
%recoverable version.